# Suppression of water vapor condensation by glycerol droplets on hydrophobic surfaces

Zhan-Long Wang[1]*, Haonan Zhao[2], Zhen Xu[1], He Hong[3]

**AFFILIATIONS**

[1]Shenzhen Institute of Advanced Technology, Chinese Academy of Sciences, Shenzhen, Guangdong 518000, China.

[2]School of Minerals Processing and Bioengineering, Central South University, Changsha, Hunan, 410000, China.

[3]Southern University of Science and Technology Hospital, Shenzhen, Guangdong 518000, China.

*Author to whom correspondence should be addressed: zl.wang1@siat.ac.cn

Abstract

'Vapor sink' is an effective strategy to suppress the formation of water vapor condensation around hygroscopic materials, and consequently reduce frosting and icing. However, traditionally used materials, such as salt solutions, fibers, exhibit insufficient condensation inhibition or pose safety concerns, such as corrosiveness. In this study, we highlight the remarkable anti-condensation properties of glycerol droplets, attributed to their strong hygroscopicity. We compared the anti-condensation capabilities of glycerol droplets with commonly used salt solutions and hygroscopic alcohols. The results indicated that glycerol droplets establish a relatively expansive dry zone where the condensation is effectively inhibited, while also offering safety compared to other materials. Furthermore, we conducted a systematic study of the anti-condensation properties of glycerol droplets with experiments and theoretical analysis. We explored the varying trends of the dry zone ratio concerning temperature, cooling time, humidity, and droplet volume on hydrophobic surfaces. To provide a comprehensive understanding, we propose a straightforward yet robust theoretical model that elucidates the relationship between this ratio and temperature, aligning well with the experimental data. Our study not only sheds light on the superior anti-condensation qualities of glycerol but also offers insights and guidelines for the development of effective anti-icing and anti-frost materials.

# 1. Introduction

Condensation, frosting and icing are fundamental natural phenomena that often lead to significant adverse issues. Examples include the traffic disruptions due to road icing, reduced power transmission efficiency caused by overhead transmission lines icing and compromised mechanical integrity, damage caused by local condensation of electronic components in precision equipment, and obstructing the external observation of the transparent cavities at low temperatures[1-6]. Therefore, the development of effective strategies to suppress the condensation and the resulting frost and ice holds immense importance in the electric industry, power harvesting, and the traffic and transportation[7-10]. Strategies to combat these problems include inhibition of ice nucleation and water vapor condensation, reduction of ice adhesion, and mitigation of frost accumulation[2,11-16]. Related researches include investigating droplet self-jumping during condensation[17], studying spontaneous droplet motion[17-20], designing and manufacturing superhydrophobic or structured surfaces[21,22], and utilizing hygroscopic materials to prevent vapor condensation[11,23]. One crucial method for suppressing frosting and icing is the inhibition of water vapor condensation of a hygroscopic liquid, often referred to as a "hygroscopic vapor sink"[11,24-27]. In recent years, scientists have designed diverse anti-frosting and anti-icing surfaces based on this concept. For example, since ice has a lower vapor pressure than liquid water, Ahmadi et al. developed a passive anti-frosting surface that employs microscopic ice stripes to suppress in-plane frost growth[28]. Sun et al. demonstrated that an array of hygroscopic droplets on solid surfaces could create extensive areas that inhibit condensation and suppress frost and ice formation[23,29]. Despite the considerable promise of 'vapor sink' in the realms of anti-icing and anti-frosting, along with the noteworthy advancements made in recent years, significant challenges and unresolved issues persist in understanding the anti-condensation properties of hygroscopic liquids and in developing effective dehumidifying materials and substances that can successfully inhibit water vapor condensation.

Hygroscopic materials possess the ability to impede the condensation of water vapor molecules in their vicinity, resulting in the formation of a dry annular area. This phenomenon arises from the reduction in humidity and subsequent decrease in dew temperature caused by the vigorous absorption of moisture during solid surface cooling. Generally, higher hygroscopicity corresponds to enhanced moisture absorption and a greater suppression of water vapor condensation. This property leads to the creation of expansive dry regions on low-temperature surfaces, playing a crucial role in mitigating condensation, particularly in the prevention of frosting and icing through the 'vapor sink' approach. Consequently, exploring effective hygroscopic materials hold great significance in anti-icing and anti-frosting. Examples of such materials include cellulose fibers[30], salts, saturated salt solutions (NaCl crystals and lithium bromide solution)[31,32], alcohols[33] (e.g., dipropylene glycol and glycerol), silica gels[34], hydrogels[12,35,36], and ice[26]. Guadarrama-Cetina et al. conducted an observation of dry zone formation around a NaCl solution droplet and examined the water vapor pressure above the droplet, revealing that the water vapor pressure above droplets of varying concentrations decreased to some extent in comparison to the saturated water vapor pressure[31,32]. Owing to osmotic pressure, hydrogels also exhibit hygroscopic properties. Delavoipière et al. measured the transfer of vapor from the atmosphere to hydrogel thin films and investigated the impact of film hygroscopy on the dynamics of film swelling[35]. Sun et al. demonstrated that glycol can effectively hinder vapor condensation, devising a bilayer-coated structure comprising a porous exterior that envelops a hygroscopic liquid-infused layer, thereby regulating vapor concentration[23,29]. Additionally, ice has also been utilized to impede vapor condensation due to its lower saturated vapor pressure relative to that of water[28]. However, these materials have various shortcomings in inhibiting condensation, such as a limited range of condensation inhibition, corrosiveness, and the need for special preparation.

Glycerol, as a cheap and commonly used industrial biosafety reagent, has a strong hygroscopicity[37-39]. However, there has been limited reports on its anti-condensation characteristics. In this study, we conducted a systematic analysis of the ability of glycerol to suppress water vapor condensation, revealing its remarkable effectiveness compared to other commonly used solutions. On a cold surface, glycerol droplets formed a conspicuously dry zone around them. We further investigated the variations in anti-condensation abilities of glycerol under varying humidity and temperature conditions, employing both experimental and theoretical approaches. Additionally, we explored the influence of droplet size and cooling time on its anti-condensation properties. To enhance our understanding of these findings, we developed a straightforward yet effective model that elucidates the relationship between the dry zone size ratio and surface temperature. Our research not only introduces a promising strategy for anti-icing and anti-frosting applications but also offers a comprehensive performance analysis and an in-depth understanding of the anti-condensation characteristics of glycerol.

## 2. Materials and methods

To demonstrate this phenomenon, we designed specialized equipment capable of capturing the dynamics of droplet condensation. This equipment comprises two main components: a capture platform for monitoring the evolution of condensation droplets and a cold source responsible for generating condensed droplets, as illustrated in Fig. 1a. The capture platform includes a CCD connected to computer, along with a corresponding optical source for imaging. The cold source section includes polydimethylsiloxane (PDMS, Dow Corning, mixed at a mass ratio of 10:1 of matrix and curing agent) thin film slice (PTFS, 2 × 2 cm$^2$ in size, 100 μm in thickness), light disk slice, semiconductor cooling chip (SCC, Model12715), DC power source, radiator and circulating water cooling pool. Experiments were conducted on a PTFS. The surface of PTFS is hydrophobic and without sharp roughness and the resulted contact line pinning[40]. To improve the observation of droplet condensation on the PTFS, we affixed the PTFS onto a light disk slice, which we cut into a 4 cm × 4 cm size. This light disk slice was placed on a SCC that is connected to a DC power source and a radiator, with a circulating water cooling pool to dissipate the heat generated during the process, as indicated in Fig. 1a. For real-time temperature monitoring, a thermocouple was securely attached to the surface of the PTFS. Temperature control was achieved by adjusting the voltage of the DC power source. By fixing the voltage, we were able to maintain a consistent temperature on the PTFS for an extended duration. To maintain controlled experimental conditions, the entire equipment setup was placed within a Temperature & Humidity Chamber (Qinzhuo, LK001), with the temperature of chamber set at 22 °C. Experiments were conducted at varying relative humidity (RH) levels, including 40% RH, 50% RH, 60% RH, and 70% RH. Glycerol was purchased from Shanghai Aladdin Biochemical Technology Co., Ltd. The contact angle of the water droplet on the surface of PTFS was approximately 120º, indicating the droplet-wise condensation on these surfaces.

The temperature variations on the PTFS are shown in Figs. 1b and 1c. In Figure 1b, we observed the temperature response when applying varying voltages to the SCC. Two points, one at the center and the other at the corner of the PTFS, were selected as illustrated in the inset of Fig. 1b. The applied voltages were 3 V, 5 V, 8 V, 10 V, and 12 V, respectively. The results reveal a rapid decrease in temperature immediately after applying power, with the system quickly reaching a stable state. This rapid cooling process typically takes less than one minute to complete. Moreover, the stabilized temperature of the PTFS can be precisely controlled by adjusting the voltage supplied by the power source. To gain deeper insights into the relationship between PDMS surface temperature and applied voltage, as well as to assess the uniformity of the surface temperature, the stabilized values for different voltages at the two selected points in were presented in Fig. 1c. The data show that the temperature initially decreases with increasing voltage and then begins to rise after passing a critical point. This behavior can be attributed to situations where the applied voltage surpasses a critical value, preventing the timely dissipation of heat generated by the SCC and thus hindering further temperature reduction on the SCC surface. The temperature difference between the two selected points is consistently small, generally remaining below 1 °C, demonstrating the relatively uniform temperature distribution across the PTFS. The adopted voltages and the corresponding temperatures in experiments are listed in Table 1. The temperatures were divided into two groups to test the condensation under different conditions.

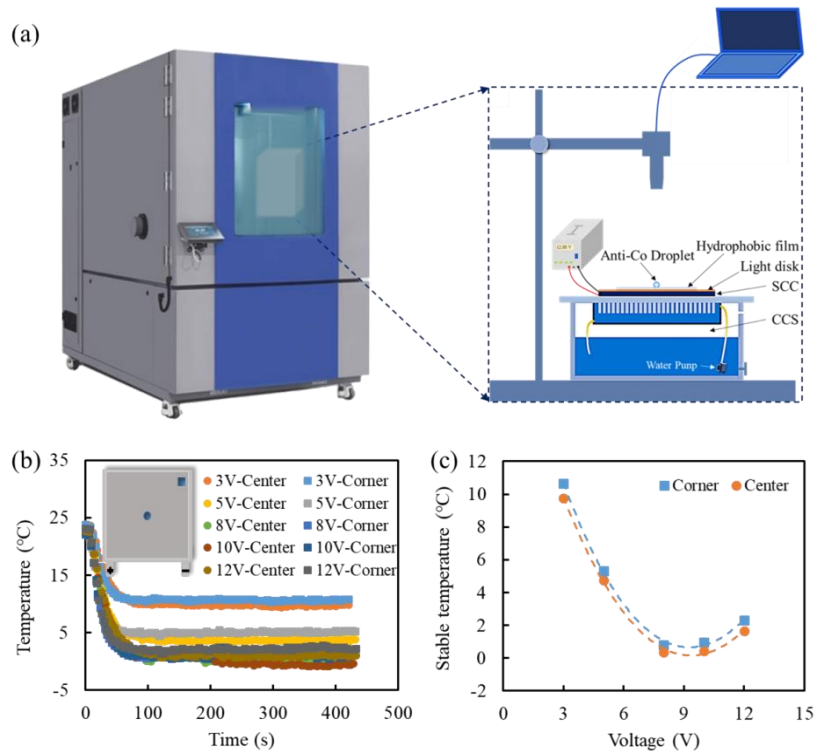

**Fig. 1.** Experimental setup and temperature control. (a) The experimental setup comprised a CCD, hydrophobic PTFS, light disk, SCC, CCS, and DC power source. The experimental setup was placed in a Temperature and Humidity Chamber to maintain a stable ambient conditions. (b) Temperature variation of PTFS after the power

source was turned on. Initially, the temperature exhibited a rapid drop, stabilizing within a span of several tens of seconds. Measurements were taken at both the center and corner of the PTFS. Insert: diagram of SCC covered by PDMS thin-film and two measurement points. (c) Variation in stabilized PTFS temperature upon voltage application.

Table 1. The adopted voltages and the corresponding temperatures in experiments

| Test conditions | | 1 | 2 | 3 | 4 | 5 | 6 |
|---|---|---|---|---|---|---|---|
| Group 1 | Voltage | 2.5 V | 3.7 V | 5.5 V | 7.5 V | 8 V | 8.8 V |
| | Temperature | 12.5 °C | 9.0 °C | 6.3 °C | 3.8 °C | 2.7 °C | 1.1 °C |
| Group 2 | Voltage | 2 V | 3 V | 4 V | 5 V | 6 V | 7 V |
| | Temperature | 13.4 °C | 11.5 °C | 8.8 °C | 6.6 °C | 6.0 °C | 4.4 °C |

The effect of glycerol hygroscopicity on the local water vapor pressure near the liquid glycerol surface was also studied, as shown in Fig. 2. The experiments were conducted within a sealed 250 ml glass container. Different volumes of glycerol liquid were added into the container. A temperature-humidity sensor was carefully positioned within the container just above the liquid glycerol. The sensor was connected to a detector produced by the Accurate Company, as illustrated in the inset of Fig. 2a. All experiments were consistently carried out with the ambient humidity level at 70%. In Fig. 2a, 100 mL of glycerol was added into the sealed container, accounting for 40% of the available space. The resulting variation in RH within the glass container after sealing was remarkable. The RH experienced a rapid decline from 70% to around 20% within a matter of seconds, following an approximately linear trend. This observation signifies the substantial capacity of glycerol to absorb moisture from the surrounding air and underscores its notable hygroscopicity. Following this abrupt decrease, the RH exhibited a more gradual decline, eventually stabilizing. During this stage, the absorption of water vapor molecules by glycerol reached a state close to dynamic equilibrium. Figure 2b shows the stabilized RH values within the glass container under various conditions, including different temperatures and glycerol contents. For a fixed volume percentage of glycerol, the experimental results showed that the RH remained almost constant at different temperatures. This indicates that the influence of temperature on the local water vapor pressure near the liquid glycerol is minimal. When we altered the volume percentage of glycerol within the container, varying RH values became evident. RH values of approximately 8%, 6%, and 2% corresponded to volume percentages of 20%, 40%, and 80% of glycerol, respectively. These results indicated that glycerol could significantly reduce the surrounding air humidity.

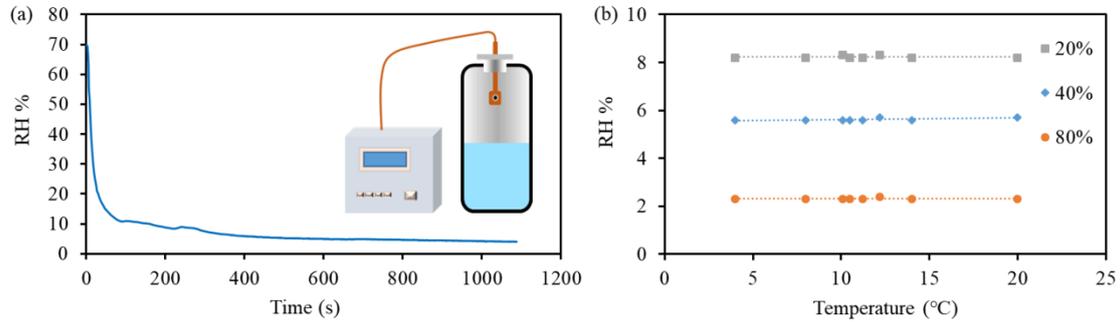

**Fig. 2.** Variation in RH near glycerol liquid under different conditions. (a) Variation in RH after sealing the glass container. Upon sealing the glass container, initially, there was a sharp decrease in RH, followed by a gradual decline. The insert shows the closed glass container alongside the humidity monitor. (b) Stabilized RH values in the closed glass container under different conditions. The results indicate that temperature exerted only a minimal influence on RH within the closed glass container. RH decreased as the volume of glycerol increased. The RH near the glycerol liquid consistently remained significantly lower than the ambient RH.

## 3. Results and discussion

### 3.1 *The performance comparison of glycerol and other generally used solutions*

The suppression of water vapor condensation could be characterized by the variations of dry zone around a glycerol droplet, which is influenced by the hygroscopicity of liquids, solid surface temperature and humidity. In a cold environment, with a fixed solid surface temperature and ambient humidity, the dry zone formation becomes more obvious as the hygroscopicity of liquids increases. Glycerol liquids have strong hygroscopicity and show excellent properties in the suppression of water vapor condensation. Furthermore, the glycerol is also low toxicity, non-corrosiveness and stable for the use of anti-condensation. To provide a comprehensive understanding, we compared the anti-condensation performance of glycerol and other commonly used and typical solutions in Fig. 3 and listed the attributes of these liquids in toxicity, corrosiveness and stability in Table 2, respectively. In the subsequent sections, we delved into the factors influencing dry zone variations, including surface temperature, cooling time, humidity, and droplet volume, offering a thorough exploration and discussion of these key parameters.

Figure 3 presents a comparative analysis of the effectiveness of eight different liquids in suppressing water vapor condensation. These liquids include glycerol, dipropylene glycol, ethylene glycol, NaCl saturated solutions, CaCl2 saturated solutions, LiBr saturated solutions, LiCl saturated solutions, and NaOH saturated solutions. In these experiments, we used a 2-μl droplet volume and maintained the PTFS surface temperature at 6.6 °C. Both vertical and oblique views of the droplets and the resulting dry zones are provided in the figures. Among the liquids tested, glycerol, $CaCl_2$ saturated solutions, LiBr saturated solutions, LiCl saturated solutions, and NaOH saturated solutions exhibited superior suppression of water vapor condensation compared to dipropylene glycol, ethylene glycol, and NaCl saturated solutions that are commonly used in anti-icing. The dry zones observed with glycerol, $CaCl_2$ saturated solutions, LiBr saturated solutions, LiCl saturated solutions, and NaOH saturated solutions were more substantial than those seen with the other liquids.

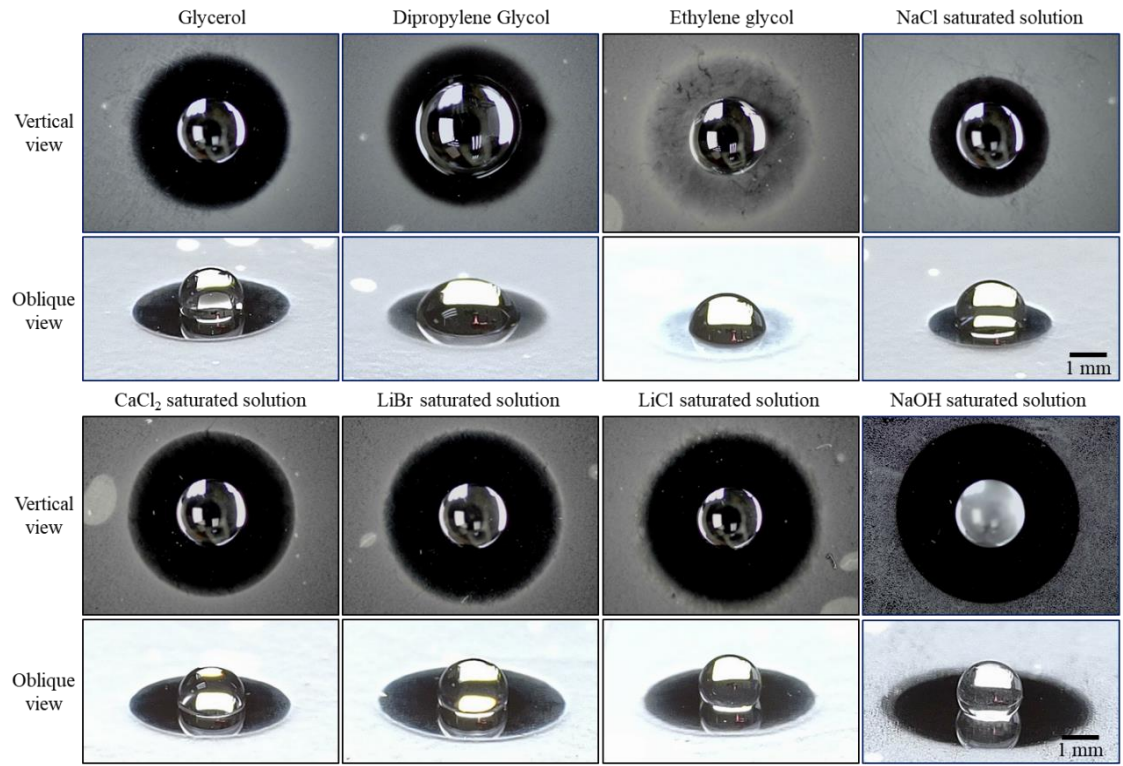

Fig. 3. Comparison of dry zone formation around hygroscopic droplets of different liquids. Eight kinds of liquids including glycerol, dipropylene glycol, ethylene glycrol, NaCl saturated solutions, CaCl2 saturated solutions, LiBr saturated solutions, LiCl saturated solutions and NaOH saturated solutions are adopted. The anti-condensation images of these droplets were observed from both top and oblique views.

In all cases, the dry zone around the NaOH saturated solutions droplet was the most extensive, as illustrated in Fig. 3. However, NaOH saturated solutions are highly corrosive, as documented in

Table 2. LiBr and LiCl are efficient water vapor absorbents and can serve as air humidity regulators in applications like absorption refrigeration. However, they also have corrosive properties that retract their application range, especially for surfaces with high requirements of corrosion resistance. $CaCl_2$, a commonly used desiccant, is employed for purposes such as road ice melting and road moisturizing. It also demonstrates excellent anti-condensation capabilities. Nevertheless, the dissolution of $CaCl_2$ in water generates a substantial amount of heat, which can pose limitations in environments demanding precise temperature control. In contrast, glycerol proves to be a highly effective agent in preventing water vapor condensation. Moreover, it possesses several advantageous properties, including low toxicity, non-corrosiveness, high water solubility, and minimal heat production upon dissolution in water. These attributes make glycerol a promising choice for various applications.

Table 2. The properties of these hygroscopic liquids

| Liquids | Suppression of condensation | Toxicity | Corrosiveness | Stability |
| --- | --- | --- | --- | --- |
| Glycerol | Strong | Low | Non-corrosiveness | Stable |
| Dipropylene glycol | Middle | Low | Non-corrosiveness | Stable |
| Ethylene glycol | Weak | Low | Non-corrosiveness | Stable |
| NaCl saturated solution | Middle | non-toxic | Non-corrosiveness | Stable |
| CaCl2 saturated solution | Strong | Low | Non-corrosiveness | Stable |
| LiBr saturated solution | Strong | non-toxic | Corrosive | Stable |
| LiCl saturated solution | Strong | Low | Corrosive | Stable |
| NaOH saturated solution | Strong | non-toxic | Strong corrosiveness | Stable |

### 3.2 *The formation of dry zone around glycerol droplets*

Figure 4 shows experimental snapshots of the variation in dry zone at different PTFS surface temperatures, with the chamber humidity controlled at 60%. The dew temperature is measured as 15.8 °C. The experiment examined two glycerol droplet volumes: 1 μl (Fig. 4a-b) and 2 μl (Fig. 4c-d). In Fig. 4a, at a temperature of 13.4 °C, close to the dew temperature, condensation occurs away from the glycerol droplet, as shown in the first image. This results in a significant dry zone around the glycerol droplet. As the temperature decreases to 11.5 °C, there is also an obvious dry zone forming a ring around the glycerol droplet. Continuing the temperature decrease from 11.5 °C to 4.4 °C, the dry zone size further diminishes. This phenomenon is attributed to the strong hygroscopic properties of the glycerol liquid, which efficiently captures water vapor molecules, leading to a reduced concentration of water vapor around the glycerol droplet. Furthermore, the RH, related to the water vapor concentration around the glycerol droplet, becomes insufficient for condensation to occur in proximity to the glycerol droplet. However, in areas farther from the glycerol droplet, the RH remains largely unaffected by the hygroscopic effect, allowing for normal condensation as the temperature decreases. Given the spherical shape of the droplet, the dry zone takes on a ring-like formation around the glycerol droplet, a mechanism further explored in the Theoretical Analysis section. The impact of temperature on the dry zone size is also illustrated in Supplementary Materials Movie S1, demonstrating a consistent reduction in dry zone size as the temperature decreases.

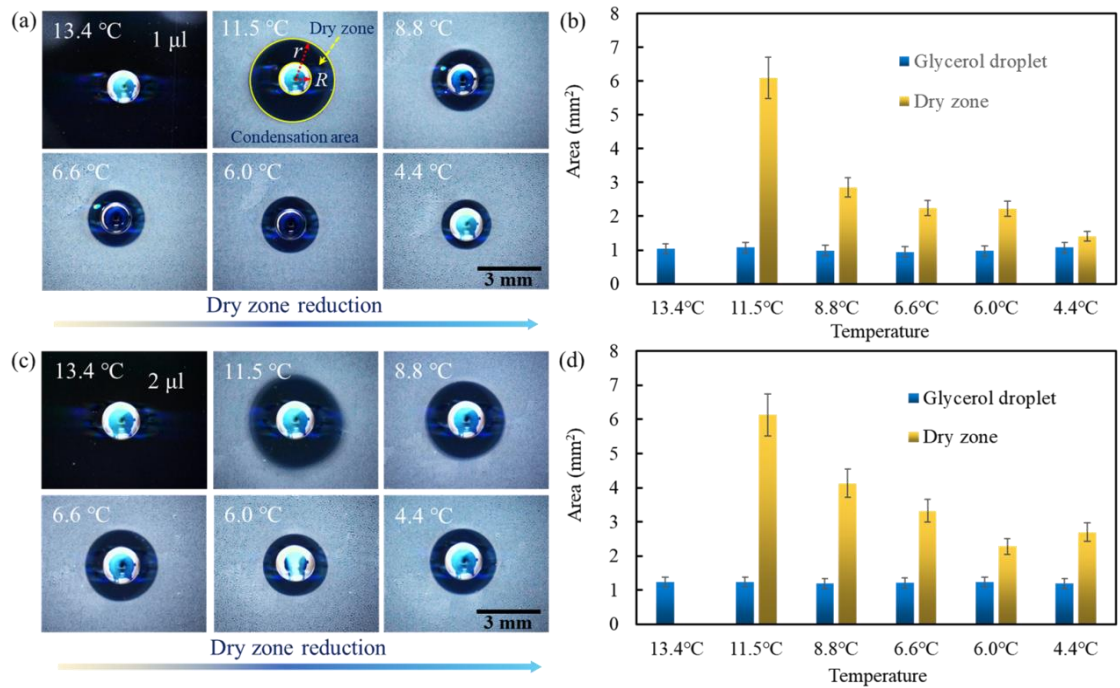

**Fig. 4.** Dry zone variation around glycerol droplet at different temperatures. Two glycerol droplet volumes are used for illustration. (a) 1-μl glycerol droplet with radius of 0.32 mm. The dry zone decreased as the substrate temperature decreased. (b) Comparison of dry zone area for 1-μl glycerol droplet. (c) 2-μl glycerol droplet with radius of 0.62

mm. The dry zone area decreased with temperature, similar to the behavior with the 1-μl droplet. (d) Comparison of dry zone area for the 2-μl glycerol droplet.

Figure 4b presents a comparison of the dry zone areas. In this figure, the blue and orange columns represent the glycerol droplet and dry zone areas, respectively. The dry zone area exhibited a significant decrease as the temperature dropped from 11.5 °C to 8.8 °C, and it continued to decrease more gradually as the temperature further decreased to 4.4 °C. Meanwhile, the area of the glycerol droplet remained relatively consistent. Figure 4c explores the scenario involving a 2-μl droplet. Compared to Fig. 4a, as the volume increases, the dry zone area also increases. However, this increase is not statistically significant. The variation in dry zone area for the 2-μl droplet follows the same trend observed for the 1-μl droplet. Figure 4d shows the areas of glycerol droplet and dry zone of 2-μl droplet. The area of dry zone decreases from 6 mm$^2$ at 11.5 °C to less than 3 mm$^2$ at 4.4 °C, showing a similar tendency with that in Fig. 4b.

The images in Fig. 5 illustrate the cases of water vapor condensation and the formation of a dry zone around glycerol droplets at different time points. The volumes of the glycerol droplets in Fig. 5 were 2 μl (Figs. 5a-b) and 3 μl (Figs. 5c-d), and the applied voltage was 4 V. The corresponding PTFS temperature was maintained at 8.8 °C throughout the experiments. In Fig. 5, the reference point '0 min' denotes the time when the power source was activated, initiating the cooling process of the PTFS. No condensation occurred around the droplet before the PTFS was cooled. After the power source was turned on, the temperature of the PTFS surface rapidly decreased, stabilizing within one minute. Thus, at 1 min, an obvious dry zone ring was formed. From 1 minute to 7 minutes, the size of dry zone remained relatively constant with a stable PTFS temperature. This phenomenon was also exhibited in Supplementary Materials Movie S2. In Figure 5b, a comprehensive comparison is provided, illustrating the changes in both the glycerol droplet and dry zone areas over time. The results reveal that the dry zone area experienced a gradual, albeit slight, reduction. This reduction can be attributed to the slow absorption of water vapor by the glycerol droplets, leading to a gradual decrease in hygroscopicity. Furthermore, the moisture absorption by the glycerol droplet resulted in a slight increase in the volume of glycerol droplet, which is demonstrated by the variations in the glycerol droplet area. The glycerol droplet volume increased with time, but the hygroscopic ability did not exhibit a significant decline within a certain time.

In Fig. 5c, the case of a 3-μl glycerol droplet is shown. The initial cooling stage is represented by 0 minute. Similar to the behavior observed with the 2-μl droplet, the dry zone showed only slight changes over time. Figure 5d shows a comparison of the glycerol droplet and dry zone areas for the 3-μl droplet. The dry zone area decreased slightly and remained almost constant for a short time. The glycerol droplet area increased slightly, but it was not noticeable. In Fig. 6, to better present the view of dry zone, the oblique view and the corresponding top view of anti-condensation of a 3-μl glycerol droplet are shown. In Fig. 6a, we provide a three-dimensional perspective that clearly shows the dry zone around the glycerol droplet, as indicated in the images in the first line. The dry zone and glycerol droplet did not undergo significant changes as exhibited in Fig. 5. Figure 6b shows the corresponding top view of the dry zone and glycerol droplet. This phenomenon indicates that the dry zone is more affected by surface temperature than by cooling time.

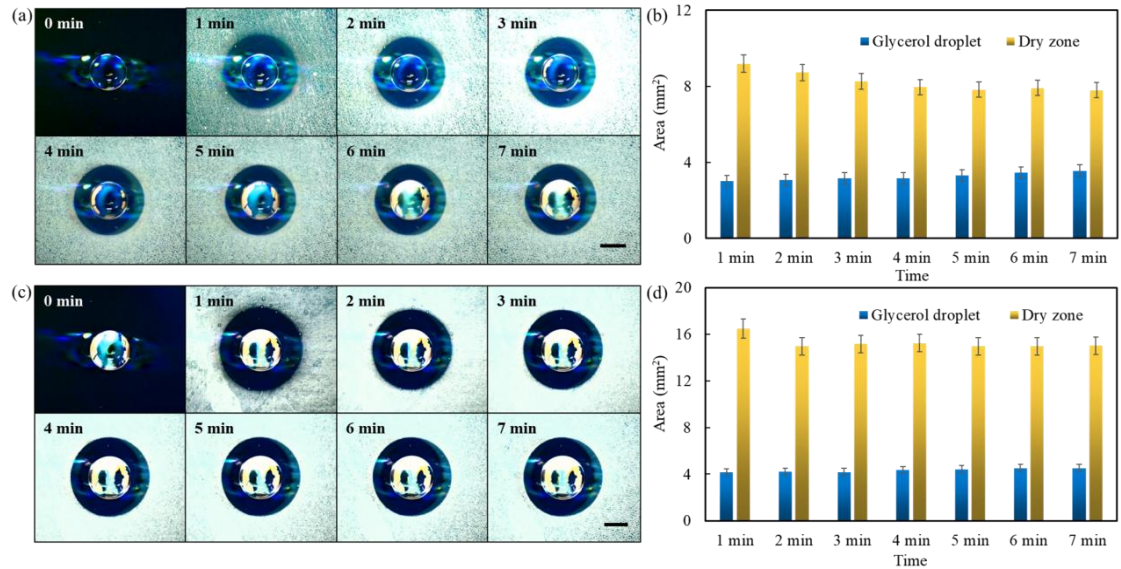

**Fig. 5.** Temporal evolution of dry zones around glycerol droplets. Two volumes are used for illustration. (a) For a 2-μl glycerol droplet, there is a minimal variation in the dry zone over time under constant ambient conditions. (b) A comparison between the glycerol droplet and dry zone areas. (c) The case of anti-condensation for a 3-μl droplet. (d) The comparison of glycerol droplet and dry zone areas of 3-μl droplet.

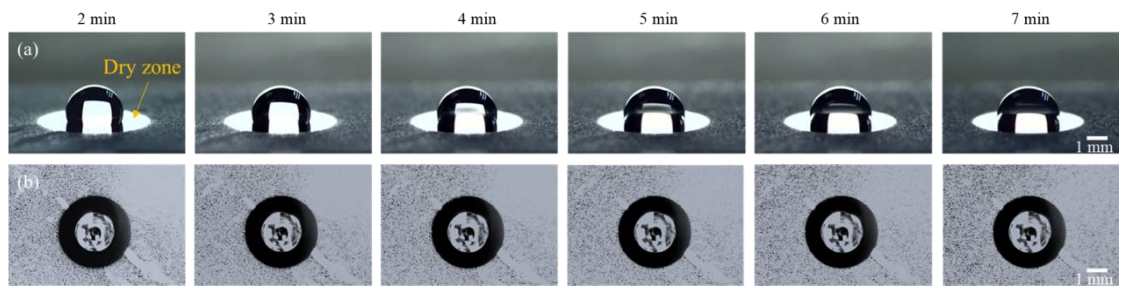

**Fig. 6.** Anti-condensation of glycerol droplet at various time points. The images depict both oblique (a) and corresponding top (b) views to offer a comprehensive view of the evolution of droplet over time.

### 3.3 *The effect of cooling time, substrate temperature, humidity and droplet volume on dry zone*

This section delves into the examination of various factors influencing the variation of the dry zone. To characterize this change, we employ a ratio, denoted as $\varsigma_{dd}$, which is defined as $\varsigma_{dd} = r/R$, where *r* represents the radius of the dry zone, and *R* stands for the radius of glycerol droplets. Figure 7 illustrates the variation of the ratio $\varsigma_{dd}$ with respect to cooling time. To establish a comprehensive understanding of this variation, we conducted experiments with different glycerol droplet volumes ranging from 1 μl to 6 μl, using two distinct groups of PTFE temperatures, referred to as Group 1 (with surface temperatures of 12.5 °C, 9.0 °C, 6.3 °C, 3.8 °C, 2.7 °C, and 1.1 °C) and Group 2 (with surface temperatures of 13.4 °C, 11.5 °C, 8.8 °C, 6.6 °C, 6.0 °C, and 4.4 °C). In Fig. 7a-f, we present the results for different cases: 1-μl droplets for Group 1 (Fig. 7a), 2-μl droplets for Group 2 (Fig. 7b), 3-μl droplets for Group 2 (Fig. 7c), 4-μl droplets for Group 1 (Fig. 7d), 5-μl droplets for Group 2 (Fig. 7e), and 6-μl droplets for Group 1 (Fig. 7f). In Fig. 7a, we observed that the ratio $\varsigma_{dd}$ remains relatively constant as the cooling time progresses under different temperature conditions, which is consistent with the findings presented in Fig. 5. This constancy can be attributed to the fact that, when the temperature remains fixed, the humidity and dew temperature around the glycerol droplet also remain relatively stable. This phenomenon further restricts the alteration of the dry zone and the surrounding condensation.

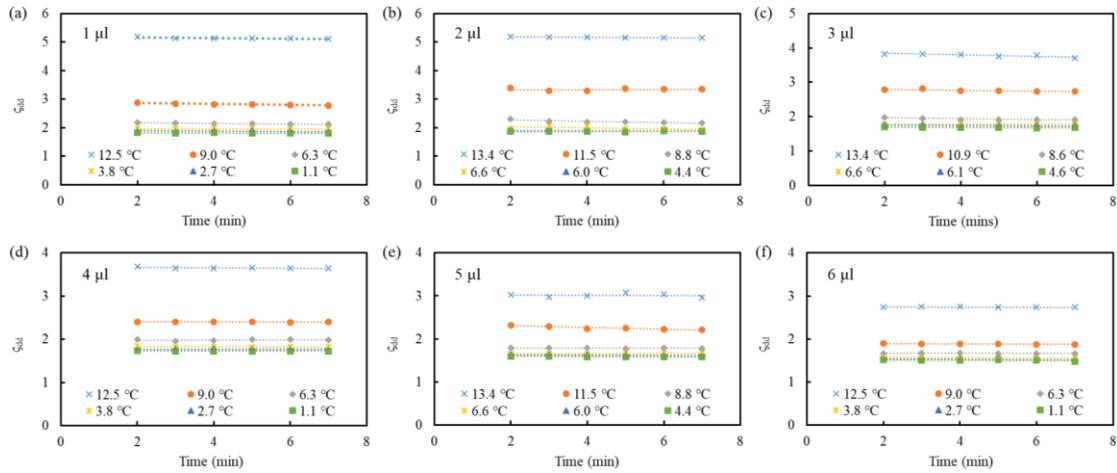

**Fig. 7.** Variation in ratio $\varsigma_{dd}$ with cooling time. The droplet volumes in (a)–(f) correspond to 1 μl (a), 2 μl (b), 3 μl (c), 4 μl (d), 5 μl (e), and 6 μl (f), respectively. The ratio $\varsigma_{dd}$ exhibits minimal change over time when subjected to consistent temperature and humidity conditions.

For glycerol droplets of different volumes but at constant temperatures, the variation in the ratio $\varsigma_{dd}$ is nearly constant during the initial stages of condensation (as seen in Figs. 7a, d, and f for Group 1), and a similar pattern is observed for Group 2. In some cases, a slight decreasing trend is discernible due to the gradual growth of glycerol droplets within moist environments. Figure 7a, for instance, illustrates that as the temperature decreases, the ratio $\varsigma_{dd}$ correspondingly decreases, indicating a clear correlation between $\varsigma_{dd}$ and temperature, a relationship elaborated in further detail in Fig. 8. Specifically, for a 1-μl droplet at 12.5 °C, $\varsigma_{dd} > 5$ (as shown in Fig. 7a), while for a 4-μl droplet (Fig. 7d), $\varsigma_{dd}$ falls to $< 4$, and for a 6-μl droplet (Fig. 7f), $\varsigma_{dd}$ drops to $< 3$. This signifies that the ratio $\varsigma_{dd}$ is inversely related to the volume of the hygroscopic droplet. Similar trends are observed for droplet volumes of 2 μl (Fig. 7b), 3 μl (Fig. 7c) and 5 μl (Fig. 7e). the change in $\varsigma_{dd}$ obeys the same rule. The maximum ratio was approximately 5, 4, and 3 for 2-μl, 3-μl, and 5-μl droplets at 13.4 °C, respectively.

The relationship between the ratio $\varsigma_{dd}$ and temperature is illustrated in Fig. 8, where different cases with varying droplet volumes are presented in Figs. 8a–f. In these figures, the dots represent the experimental data, while the dotted lines depict the fitting curves. The ratio $\varsigma_{dd}$ was recorded at various time points, and notably, it remained relatively consistent across different times, agreeing with the findings displayed in Fig. 7. For these experiments, the room temperature and RH in the chamber were maintained at 22°C and 60%, respectively. Due to fluctuations in airflow, the measured dew temperatures varied slightly among experiments in Fig. 8: 15.3°C (Fig. 8a), 15.7°C (Fig. 8b), 15.5°C (Fig. 8c), 15.6°C (Fig. 8d), 15.8°C (Fig. 8e), and 15.4°C (Fig. 8f), all falling within the range of 15.5±0.3°C. In Fig. 8a, we examined the variation in the ratio $\varsigma_{dd}$ for a 1-μl droplet. The results reveal a direct relationship between $\varsigma_{dd}$ and temperature, with the ratio increasing as the temperature rises and approaching infinity as it approaches the dew temperature. Conversely, as the temperature decreases, the ratio $\varsigma_{dd}$ and the area of dry zone diminishes. This change in $\varsigma_{dd}$ with temperature follows a hyperbolic function. It is also discussed with a theoretical model in the following section of Theoretical Analysis. In Fig. 8b, we expand our investigation to a 2-μl glycerol droplet, and the data exhibit a similar trend as observed in Fig. 8a. This consistent trend in Figures 8c–f further reinforces the pattern.

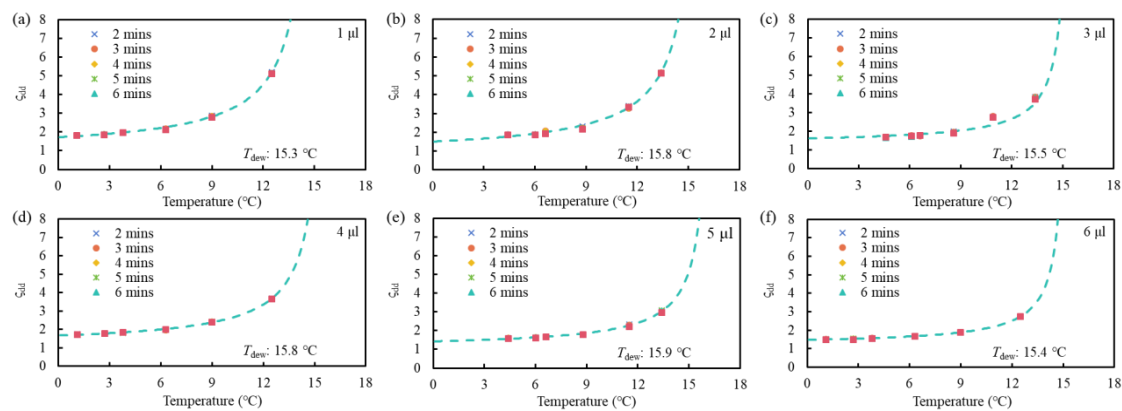

**Fig. 8.** Variation in ratio $\varsigma_{dd}$ with substrate temperature. Different glycerol droplet volumes are used for illustration: in (a)–(f), the volumes are 1 μl (a), 2 μl (b), 3 μl (c), 4 μl (d), 5 μl (e), and 6 μl (f), respectively. The dots represent the experimental data at different times and temperatures. The dotted line represent the fitting curve.

The effect of humidity on the ratio $\varsigma_{dd}$ was examined at RH of 40%, 50%, 60%, and 70%, as depicted in Fig. 9. Different temperatures and droplet volumes ranging from 1 to 6 μl in 1 μl increments were employed. Figure 9a illustrates the scenario at 40% RH, with measurements taken at temperatures within Group 2. The change in $\varsigma_{dd}$ with PTFS temperature adhered to the same pattern observed in Fig. 8. For different droplet volumes, the variation in $\varsigma_{dd}$ exhibited a consistent trend. Notably, in Fig. 9a, the highest ratio, approximately 5.72, was observed for a 1-μl droplet, surpassing that of other volumes at the same temperature. Conversely, the smallest ratio, 1.56, was recorded for a 6-μl droplet at 1.8 °C. This indicates a discernible relationship: larger volumes corresponded to smaller ratios. In Fig. 9b, the RH is 50%. The maximum $\varsigma_{dd}$ is 5.19 for a 1-μl droplet at 12.2 °C, slightly lower than that at 40% RH. On the other hand, the minimum $\varsigma_{dd}$ was 1.51 for a 6-μl droplet at 1.1 °C. In Fig. 9c, the RH is 60%. The maximum $\varsigma_{dd}$ was 4.4 for a 1-μl droplet at 14.7 °C. The minimum $\varsigma_{dd}$ was 1.51 at 1.6 °C. In Fig. 9d, the RH is 70%. The maximum $\varsigma_{dd}$ was 3.77 for a 1-μl droplet at 14.7 °C, smaller than in all other cases. The minimum $\varsigma_{dd}$ was 1.38 at 2.0 °C for a 6-μl droplet. The results indicate a decrease in $\varsigma_{dd}$ with increasing humidity under the same conditions. Furthermore, the disparity in $\varsigma_{dd}$ between droplets of varying volumes tended to diminish as RH increased. This can be attributed to the heightened humidity surrounding the glycerol droplets. When the substrate temperature was held constant, higher RH facilitated easier vapor condensation, leading to a further reduction in the dry zone area. Consequently, high RH levels led to a decrease in $\varsigma_{dd}$.

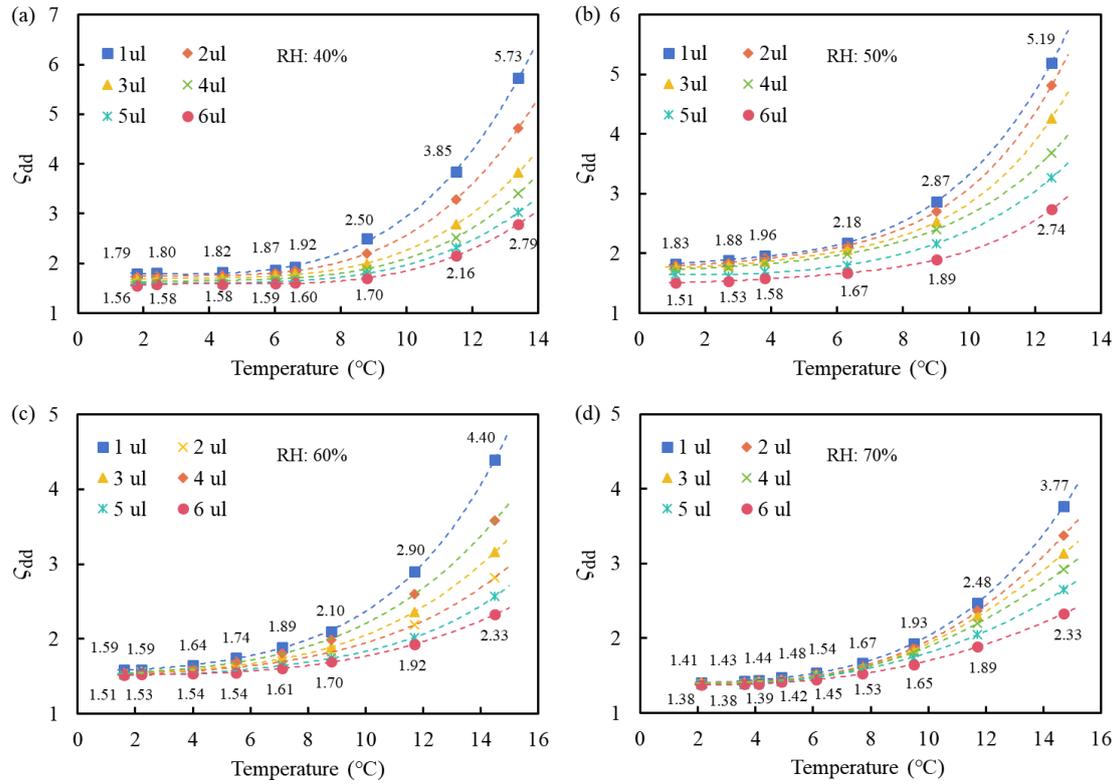

**Fig. 9.** Variation in ratio $\varsigma_{dd}$ under various RH conditions. (a) RH = 40%. (b) RH = 50%. (c) RH = 60%. (d) RH = 70%. The results indicate that as the RH increases, the difference in $\varsigma_{dd}$ between cases with different droplet volumes becomes less pronounced.

Figure 10 illustrates the impact of glycerol droplet volume on the $\varsigma_{dd}$ ratio. It is evident from the graph that $\varsigma_{dd}$ decreases as the glycerol droplet volume increases. The data in Fig. 10 are categorized into two temperature groups. In Fig. 10a, the temperatures of Group 1 are adopted. The results showed that $\varsigma_{dd}$ decreased with the glycerol droplet volume, especially at high temperatures. However, at lower temperatures, $\varsigma_{dd}$ exhibited only minor variations, remaining almost constant across different glycerol droplet volumes. In Fig. 10a, $\varsigma_{dd}$ ranged from 5.2 to 2.7 at 12.5°C and from 1.8 to 1.5 at 1.1°C for droplet volumes of 1 μl, 4 μl, and 6 μl. Figure 10b shows cases with temperatures of Group 2. The variation in $\varsigma_{dd}$ was similar to that shown in Fig. 10a. The maximum $\varsigma_{dd}$ ranged from 5.1 to 3.0 at 13.4 °C for 2-μl, 3-μl, and 5-μl droplets. The minimum $\varsigma_{dd}$ ranged from 1.8 to 1.6.

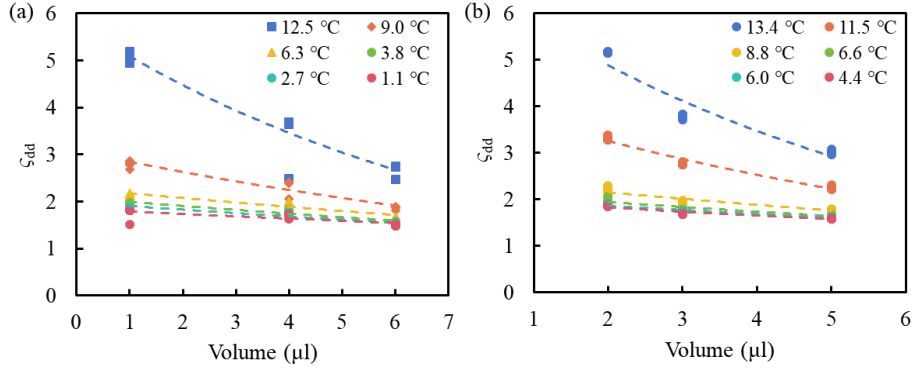

**Fig. 10.** Change in $\varsigma_{dd}$ with glycerol droplet volume for two temperature groups. (a) 1-μl, 4-μl, and 6-μl droplets at temperatures of Group 1. The ratio $\varsigma_{dd}$ decreased with the droplet volume under the same temperature conditions. (b) 2-μl, 3-μl, and 5-μl droplets the temperatures of Group 2. Results show a similar tendency with (a).

## 4. Theoretical analysis

A theoretical model was introduced to determine the mechanism controlling the evolution of the dry zone. In our theoretical framework, we considered an immobile droplet with a radius $R$ placed on a horizontal plane, as depicted in Fig. 11a. To facilitate our analysis, we defined the horizontal and vertical planes as the $H$ and $V$ planes, respectively. Water vapor molecules (WVMs) were diffused around the droplets. The WVM concentration $n(r, t)$ obeyed the following equation:

$$\frac{\partial n}{\partial t} = D_0 \Delta n, \tag{1}$$

where, $D_0$ and $t$ denote the diffusion coefficient and time, respectively. $r$ is the distance to the center of the droplet. The hydroscopic droplet radius evolution follows the growth equation:

$$\frac{dV}{dt} = \int J dS. \tag{2}$$

where, $J$ is the flux of the WVMs per unit surface area, and $V$ is the droplet volume. The WVM flux can be expressed as:

$$J = D_0 \left. \frac{\partial n}{\partial r} \right|_R. \tag{3}$$

The evolution of the droplet and dynamic dry zone is a Stefan problem with a moving boundary at $r = R(t)$. We first discuss the influence of temperature on the variation in the dry zone. We assume sufficiently slow growth, as shown in Eq. (1), such that the time dependence of $n$ can be neglected (quasistatic approximation). We obtain the Laplace equation:

$$\Delta n = 0. \tag{4}$$

For the 3D space, Eq. (4) has a hyperbolic solution[31]:

$$n = n_\infty - (n_\infty - n_R)\frac{R}{r}, \tag{5}$$

The boundary conditions are $n(R) = n_R$ and $n(\infty) = n_\infty$, where $n_R$ is the WVM concentration at the surface of the hygroscopic droplet.

For the dry zone evolution (the anti-condensation properties), we considered the cooling substrate surface, the surface where condensation occurred (here, the $H$ plane), as shown in Fig. 11b. In this case, the distance $r$ is the length from the center of the hygroscopic droplet to the edge of the dry zone. For the variation in the dry zone size, the distance $r$ was determined through comparison of the substrate surface temperature and dew temperature $T_{dew}$. The critical condition determining the edge of the dry zone was that the cooling surface temperature $T_c$ was equal to the dew temperature $T_{dew}$ at the edge of the dry zone,

$$T_c = T_{dew}. \tag{6}$$

In general, in certain atmospheric environments, $T_{dew}$ can be calculated using the following equation, summarized by Hyland and Wexter[41]:

$$T_{dew} = \varphi(A + B \cdot T_c) + C \cdot T_c - 19.2, \tag{7}$$

where, A, B, and C are constant parameters with values of 0.1980, 0.0017, and 0.8400, respectively[41]. The WVM concentration $n$ can be characterized by the moisture content $d$. The moisture content is related to $\varphi$, water vapor pressure $P_q$, saturated water vapor pressure $P_s$ ($P_q = \varphi P_s$), and local air pressure $P$ as follows:

$$n = d = 622\frac{\varphi P_s}{P - \varphi P_s}. \tag{8}$$

Integrating, we obtain:

$$\frac{\varphi P_s}{P - \varphi P_s} = \frac{P_s}{P - P_s} - \left(\frac{P_s}{P - P_s} - \frac{\varphi_G P_s}{P - \varphi_G P_s}\right) \cdot \frac{1}{\varsigma_{dd}}, \tag{9}$$

$$\varphi(A + B \cdot T_c) + C \cdot T_c - 19.2 = T_c, \tag{10}$$

where, $\varphi_G$ is the RH at the top of the hygroscopic droplets. $P$ is generally regarded as the atmospheric pressure. $P_s$ was obtained from an atmospheric handbook. By solving Eqs. (9) and (10), the relationship between the ratio $\varsigma_{dd}$ and the PTFS temperature $T_c$ can be obtained. In Fig. 11b, the RH in the dry zone is lower than that in the condensation area, i.e., RH<RH$_{de}$, where RH$_{de}$ is the critical RH at the dew temperature. In the condensation area, RH is greater than RH$_{de}$. Figure 11c shows the relationship between $T_{dew}$ and RH at room temperature (22 °C and 25 °C)[41]. The plot indicates that the dew temperature changes linearly with the RH in a stable atmospheric environment.

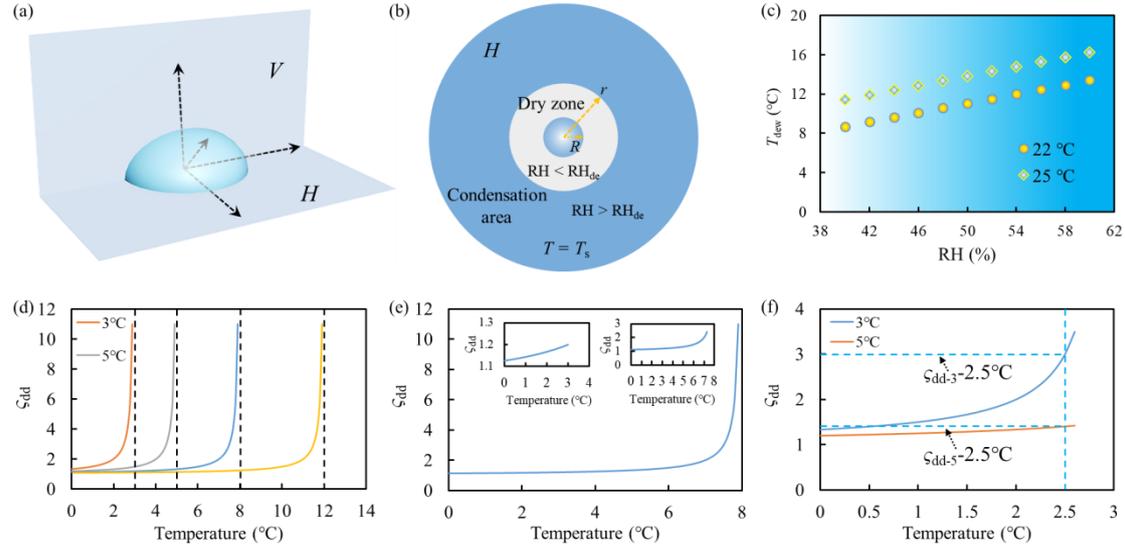

**Fig. 11.** Diagram of hydroscopic droplet and dry zone, and theoretical analysis results. (a) Diagram of hygroscopic droplet. $H$ represents horizontal plane, $V$ represents vertical plane. (b) Diagram of dry zone in $H$ plane. RH is lower than $RH_{de}$ in the dry zone but higher than $RH_{de}$ in the condensation area. (c) Dew temperature at different RH levels. (d) The variation of ratio $\varsigma_{dd}$ at different temperatures, especially at $T_{dew}$ of 3 °C, 5 °C, 8 °C, and 12 °C. (e) Change tendency of ratio $\varsigma_{dd}$ in different parts of the temperature period. (f) Comparison of ratio $\varsigma_{dd}$ at $T_{dew}$ = 3 °C and $T_{dew}$ = 5 °C under the same temperature conditions.

Figures 11d–e show the theoretical curves between the ratio $\varsigma_{dd}$ and the substrate surface temperature. In Fig. 11d, four $T_{dew}$ values are shown (3 °C, 5 °C, 8 °C, and 12 °C). From the theoretical analysis and curves, the variation in the ratio $\varsigma_{dd}$ with the substrate surface temperature indicates a hyperbolic function. As the temperature approaches the dew temperature from a lower value, the ratio steadily increases, ultimately approaching infinity as it nears $T_{dew}$. Conversely, as the temperature decreased, the ratio converges toward a specific value, which is determined by the properties of the hygroscopic liquid. In the range from 0 °C to $T_{dew}$, in most ranges, the ratio $\varsigma_{dd}$ exhibits gradual increments; as it approaches $T_{dew}$, it sharply escalates. For different values of $T_{dew}$, a similar rule for the ratio variation was observed. However, when we consider the temperature range significantly distant from $T_{dew}$, the variation in the ratio $\varsigma_{dd}$ suggests an approximately linear relationship, as illustrated in Fig. 11e. In Fig. 11e, the case of $T_{dew}$ = 8 °C is shown. The inset images provide a detailed view of the ratio variations across different temperature intervals. The first inset picture pertains to the temperature range from 0 °C to 3 °C, and it exhibits an approximately linear

change. In the second inset picture, encompassing the temperature range from 0 °C to 7 °C, the curve initially changes smoothly and subsequently enters a nonlinear increasing phase. Figure 11f compares the ratio of $\varsigma_{dd-3}$ and $\varsigma_{dd-5}$ (corresponding to $T_{dew}$ values of 3 °C and 5 °C) at 2.5 °C. From the curves, a noticeable distinction between the two cases at 2.5 °C can be observed. The ratio is close to 3 when $T_{dew}$ = 3 °C and slightly exceeds 1.2 when $T_{dew}$ = 5 °C. This stark contrast highlights the significant influence of the dew temperature on the variation in the dry zone.

In Fig. 12, a comparison between experimental data and a theoretical model is presented. Six distinct cases, characterized by different volumes and temperature groups, are exhibited in Fig. 12a through Fig. 12f. These cases are as follows: 1 μl-Group 1 (Fig. 12a), 2 μl-Group 2 (Fig. 12b), 3 μl-Group 2 (Fig. 12c), 4 μl-Group 1 (Fig. 12d), 5 μl-Group 2 (Fig. 12e), and 6 μl-Group 1 (Fig. 12f). In these figures, dots represent the experimental data, while the dotted line corresponds to the theoretical model. The relationship between the ratio $\varsigma_{dd}$ and $1/(T_{dew}-T_c)$ is depicted. It is evident from the results that the experimental data align closely with the theoretical curve. This alignment underscores the effectiveness of our proposed model in accurately describing and predicting changes in dry zones.

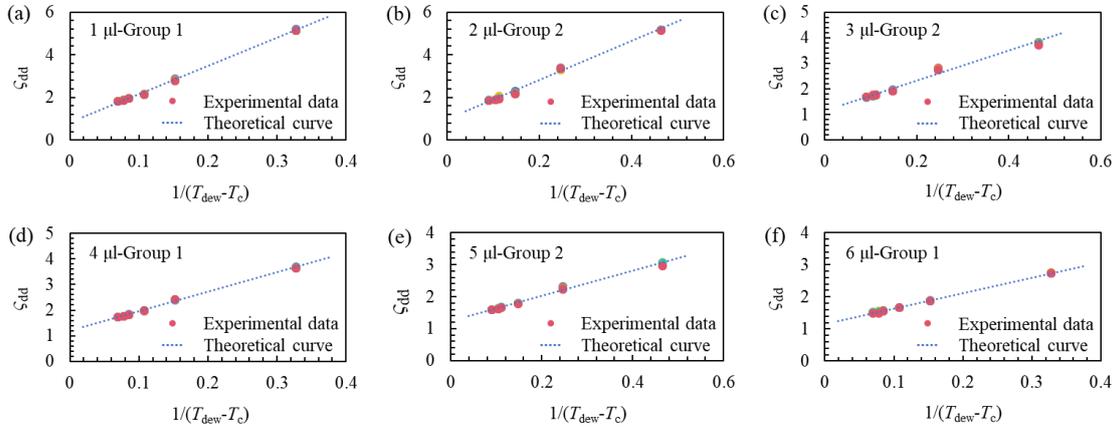

**Fig. 12.** The comparison of experimental data and the theoretical model. (a)-(f) showed different cases of experimental data and theoretical curve: 1 μl droplet and Group 1 (a), 2 μl droplet and Group 2 (b), 3 μl droplet and Group 2 (c), 4 μl droplet and Group 1 (d), 5 μl droplet and Group 2 (e) and 6 μl droplet and Group 1 (f). The dots represent experimental data. The dotted line represents the theoretical curve. The theoretical curve agrees with the experimental data well.

## 5. Conclusions

In this study, the anti-condensation properties of glycerol droplets were studied experimentally and theoretically. To control the temperature of the PTFS, a device combining an SCC and CCS was employed. All experiments were conducted inside a Temperature & Humidity Chamber, which ensured a stable environment for controlling temperature and humidity conditions. The tests on the hygroscopicity of glycerol liquid showed that glycerol liquid could significantly reduce the humidity in a local space near it. Compared with other hygroscopic solutions, glycerol has excellent performance in suppressing water vapor condensation, while also possessing characteristics of low toxicity, non-corrosiveness, and stability. The variation of dry zone around glycerol droplet was analyzed and discussed with experiments and theoretical analysis. The results showed that the dry zone area increased with the PTFS surface temperature and droplet volume. As the temperature approached the dew temperature, the area of the dry zone approached infinity. To quantify the anti-condensation properties of glycerol droplets, a parameter $\varsigma_{dd}$, defined as the ratio of the dry zone radius to the glycerol droplet radius, was proposed to characterize the anti-condensation properties of glycerol droplets. The effects of the PTFS temperature, cooling time, RH, and glycerol droplet volume were considered. The results showed a hyperbolic relationship between $\varsigma_{dd}$ and temperature, whereas it remained relatively constant with varying cooling times and decreased with rising humidity and glycerol droplet volume. To provide a comprehensive understanding of this relationship, we developed a straightforward yet effective model to illustrate how $\varsigma_{dd}$ varies with surface temperature. The theoretical curves closely matched the experimental data, confirming the accuracy of our model. This study provides an intuitive view of the anti-condensation properties of glycerol droplets, and offer a basis for design of anti-condensation materials using the 'vapor sink' method.

Acknowledgments:

This work was supported by the National Natural Science Foundation of China (Nos. 12202461, 31800871), Shenzhen Science and Technology Research Program (JCYJ20210324101603009).